\documentclass[10pt,conference,]{IEEEtran}
\usepackage{cite}
\usepackage{amsmath,amssymb,amsfonts}
\usepackage{algorithmic}
\usepackage{graphicx}
\usepackage{textcomp}
\usepackage{xcolor}
\usepackage{balance}

\def\BibTeX{{\rm B\kern-.05em{\sc i\kern-.025em b}\kern-.08em
    T\kern-.1667em\lower.7ex\hbox{E}\kern-.125emX}}

\usepackage{defs}
\graphicspath{{./}{./img/}}
\usepackage{paralist}
\usepackage{xfrac}

\usepackage{tikz}
\usetikzlibrary{arrows}
\usetikzlibrary{chains}
\usetikzlibrary{fadings}
\usetikzlibrary{matrix}
\usetikzlibrary{fit}
\usetikzlibrary{positioning}
\usetikzlibrary{shapes}
\usetikzlibrary{automata}
\tikzset{
	ctrlstate/.style = {state,align=center,inner sep=2pt, minimum size=2mm},
	cfastate/.style = {ctrlstate},
	cfatargetstate/.style = {cfastate, double},
	compstate/.style = {cfastate, rounded rectangle, minimum height=5mm, minimum width=3em, inner sep=3pt},
	concept/.style = {cfastate, inner sep=3pt, fill=gray!50, rectangle,
		minimum width=17mm, minimum height=9mm, draw=gray!6 },
	crosscutting/.style = {concept, rounded rectangle, fill=gray!30},
	conceptstate/.style = {concept, rounded rectangle, fill=gray!30, draw=gray!90, minimum width=20mm},
	inputstate/.style = {concept, rectangle, fill=gray!10, draw=gray!90, minimum width=20mm, inner sep=3pt},
	explain/.style = {circle, draw=gray!30, line width=1mm, minimum size=7mm},
	one/.style = {fill=blue!70,draw=blue!70},
	two/.style = {fill=red!50,draw=red!50},
	abststate/.style = {rectangle,align=center,inner sep=2pt,minimum size=3.5mm,fill=gray!0, draw=gray!90},
	line/.style = {draw},
	trans/.style = {draw,semithick,->,shorten >=1pt,>=stealth'},
	missing/.style = {draw=red,densely dotted,fill=red,semithick,->,shorten >=1pt,>=stealth'},
	ctrans/.style = {draw,very thick,->,shorten >=1pt,>=stealth',draw=gray!90},
	epsilon/.style = {trans,dashed},
	strengthen/.style = {draw=gray!30,semithick,double,shorten >=1pt,>=stealth',line width=1mm},
}

\newcommand{\motionblockleft}{\begin{mbox}\sf\begin{tikz}[baseline=(X.base)]\node[draw=black!60,fill=blue!12,semithick,rectangle,inner sep=1pt, minimum size=1em, outer sep=0pt, rounded corners=1pt] (X)}%
\newcommand{\controlblockleft}{\begin{mbox}\sf\begin{tikz}[baseline=(X.base)]\node[draw=black!60,fill=orange!15,semithick,rectangle,inner sep=1pt, minimum size=1em, outer sep=0pt, rounded corners=1pt] (X)}%
\newcommand{\hatblockleft}{\begin{mbox}\sf\begin{tikz}[baseline=(X.base)]\node[draw=black!60,fill=yellow!20,semithick,rectangle,inner sep=1pt, minimum size=1em, outer sep=0pt, rounded corners=1pt] (X)}%
\newcommand{\looksblockleft}{\begin{mbox}\sf\begin{tikz}[baseline=(X.base)]\node[draw=black!60,fill=violet!20,semithick,rectangle,inner sep=1pt, minimum size=1em, outer sep=0pt, rounded corners=1pt] (X)}%
\newcommand{\sensingblockleft}{\begin{mbox}\sf\begin{tikz}[baseline=(X.base)]\node[draw=black!60,fill=cyan!20,semithick,rectangle,inner sep=1pt, minimum size=1em, outer sep=0pt, rounded corners=1pt] (X)}%
\newcommand{\soundblockleft}{\begin{mbox}\sf\begin{tikz}[baseline=(X.base)]\node[draw=black!60,fill=magenta!20,semithick,rectangle,inner sep=1pt, minimum size=1em, outer sep=0pt, rounded corners=1pt] (X)}%
\newcommand{\operatorblockleft}{\begin{mbox}\sf\begin{tikz}[baseline=(X.base)]\node[draw=black!60,fill=green!20,semithick,rectangle,inner sep=1pt, minimum size=1em, outer sep=0pt, rounded corners=1pt] (X)}%

\newcommand{\blockleft}{\begin{mbox}\sf\begin{tikz}[baseline=(X.base)]\node[draw=black!60,fill=black!3,semithick,rectangle,inner sep=1pt, minimum size=1em, outer sep=0pt, rounded corners=1pt] (X)}%
\newcommand{\blockright}{;\end{tikz}\normalfont\end{mbox}}%

\newcommand{\mblockleft}{\begin{mbox}\sf\begin{tikz}[baseline=(X.base)]\node[draw=red!60,densely dotted,fill=red!3,semithick,rectangle,inner sep=1pt, minimum size=1em, outer sep=0pt, rounded corners=1pt] (X)}%
\newcommand{\mblockright}{;\end{tikz}\normalfont\end{mbox}}%

\newcommand\accOneOne{\SI[round-mode=places, round-precision=2]{6.244796618602612}{\percent}\xspace}
\newcommand\accOneTwo{\SI[round-mode=places, round-precision=2]{12.271605424817604}{\percent}\xspace}
\newcommand\accOneThree{\SI[round-mode=places, round-precision=2]{17.1726364948545}{\percent}\xspace}
\newcommand\accOneFive{\SI[round-mode=places, round-precision=2]{25.875522801278862}{\percent}\xspace}
\newcommand\accOneTen{\SI[round-mode=places, round-precision=2]{44.378993364303206}{\percent}\xspace}
\newcommand\accTwoOne{\SI[round-mode=places, round-precision=2]{23.223214593606674}{\percent}\xspace}
\newcommand\accTwoTwo{\SI[round-mode=places, round-precision=2]{34.55611769863985}{\percent}\xspace}
\newcommand\accTwoThree{\SI[round-mode=places, round-precision=2]{42.54247683417655}{\percent}\xspace}
\newcommand\accTwoFive{\SI[round-mode=places, round-precision=2]{54.03117348874099}{\percent}\xspace}
\newcommand\accTwoTen{\SI[round-mode=places, round-precision=2]{69.4413107841157}{\percent}\xspace}
\newcommand\accThreeOne{\SI[round-mode=places, round-precision=2]{31.410812195494426}{\percent}\xspace}
\newcommand\accThreeTwo{\SI[round-mode=places, round-precision=2]{43.87072460626721}{\percent}\xspace}
\newcommand\accThreeThree{\SI[round-mode=places, round-precision=2]{52.08012098939372}{\percent}\xspace}
\newcommand\accThreeFive{\SI[round-mode=places, round-precision=2]{62.82458015793648}{\percent}\xspace}
\newcommand\accThreeTen{\SI[round-mode=places, round-precision=2]{76.07405181459458}{\percent}\xspace}
\newcommand\accFourOne{\SI[round-mode=places, round-precision=2]{36.306054888592215}{\percent}\xspace}
\newcommand\accFourTwo{\SI[round-mode=places, round-precision=2]{49.04528727591592}{\percent}\xspace}
\newcommand\accFourThree{\SI[round-mode=places, round-precision=2]{56.77030735049977}{\percent}\xspace}
\newcommand\accFourFive{\SI[round-mode=places, round-precision=2]{66.75513441350195}{\percent}\xspace}
\newcommand\accFourTen{\SI[round-mode=places, round-precision=2]{78.35147024774254}{\percent}\xspace}

\newcommand\bugramOutlierFour{\SI[round-mode=places, round-precision=2]{27.27}{\percent}\xspace}

\newcommand\bugramOutlierFive{\SI[round-mode=places, round-precision=2]{90.91}{\percent}\xspace}

\newcommand\accTransOne{\SI[round-mode=places, round-precision=2]{33.826082000424873}{\percent}\xspace}
\newcommand\accTransTwo{\SI[round-mode=places, round-precision=2]{43.782998875641843}{\percent}\xspace}
\newcommand\accTransThree{\SI[round-mode=places, round-precision=2]{49.905180855859355}{\percent}\xspace}
\newcommand\accTransFive{\SI[round-mode=places, round-precision=2]{57.79356369722123}{\percent}\xspace}
\newcommand\accTransTen{\SI[round-mode=places, round-precision=2]{69.32237990870419}{\percent}\xspace}

\newcommand\accTransControl{0.3103623256030983}
\newcommand\accTransData{0.66275682222392}
\newcommand\accTransEvent{0.30413377769981875}
\newcommand\accTransLooks{0.6417587912774811}
\newcommand\accTransMotion{0.48867989541317874}
\newcommand\accTransPen{0.2511453561016243}
\newcommand\accTransSensing{0.7587106606017225}
\newcommand\accTransSound{0.4471932439145554}
\newcommand\accTransOperator{0.5323626495492322}
\newcommand\accTransMyblocks{0.605557750559726}

\newcommand\accTransShapeEnd{0.5889020303567909}
\newcommand\accTransShapeHat{0.16926518438177873}
\newcommand\accTransShapeStack{0.562980762394961}
\newcommand\accTransShapeC{0.14034484502199523}
\newcommand\accTransShapeDiamond{0.7065991183630502}
\newcommand\accTransShapeOval{0.6759186179075681}

\newcommand{\accTransEventBackdropSwitch}{35.2941176470588}
\newcommand{\accTransEventKeyPressed}{25.2780704835499}
\newcommand{\accTransEventBroadcaseRecv}{22.4796096979628}

\newcommand{\pvalueBlocks}{0.191\xspace}

\begin{document}

\title{On the Applicability of Language Models\\ to Block-Based Programs}

\author{%
  \IEEEauthorblockN{Elisabeth Griebl\IEEEauthorrefmark{1}, Benedikt Fein\IEEEauthorrefmark{1}, Florian Obermüller\IEEEauthorrefmark{1}, Gordon Fraser\IEEEauthorrefmark{1}, René Just\IEEEauthorrefmark{2}}
  \IEEEauthorblockA{\IEEEauthorrefmark{1}\textit{University of Passau}, Passau, Germany}
  \IEEEauthorblockA{\IEEEauthorrefmark{2}\textit{University of Washington}, Seattle, USA}
  \IEEEauthorblockA{\{elisabeth.griebl, benedikt.fein, florian.obermueller, gordon.fraser\}@uni-passau.de, rjust@cs.washington.edu}
}

\maketitle

\begin{abstract}
  %
  Block-based programming languages like \Scratch
  are increasingly popular
  for programming education and end-user programming.
  %
  Recent program analyses build on the insight that source code can be modelled
  using techniques from natural language processing.
  %
  Many of the regularities of source code that support this approach are due to
  the syntactic overhead imposed by textual programming languages. This
  syntactic overhead, however, is precisely what block-based languages remove
  in order to simplify programming. Consequently, it is unclear how well this
  modelling approach performs on block-based programming languages.
  %
  In this paper, we investigate the applicability of language models
  for the popular block-based programming language \Scratch. We model
  \Scratch programs using \mbox{n-gram} models, the most essential type of
  language model, and transformers, a popular deep learning model.
  Evaluation on the example tasks of code completion and bug finding
  confirm that blocks inhibit predictability, but the use of language
  models is nevertheless feasible.
  %
  %
  Our findings serve as foundation for improving tooling and analyses for
  block-based languages.
  %
\end{abstract}


\section{Introduction}\label{sec:introduction}


Block-based programming languages are becoming increasingly popular
for education~\cite{mcgill2020} as well industrial
applications requiring end-user
programming~\cite{weintrop2017blockly,ritschel2020comparing,mayr2021considerations}. The
distinguishing feature of these programming languages is that they
reduce the syntactic overhead that is common for text-based languages,
and instead represent programming constructs using graphical blocks
which can only be combined in syntactically valid
ways. \Cref{fig:example:java} shows a \java function that
prints \enquote{Hello world!} 10 times; the same functionality can be
implemented in the popular block-based language
\Scratch~\cite{maloney2010} with only three blocks
(\cref{fig:example:scratch}).  Programming is usually further
simplified by explicitly listing all available blocks in the user
interface, such that programmers neither need to memorize syntax nor
available commands and APIs (recognition over recall).

\newsavebox{\firstlisting}
\begin{lrbox}{\firstlisting}
\begin{minipage}[b]{0.61\columnwidth}
\begin{footnotesize}
\begin{lstlisting}[frame=None,language=Java]
public static void main(
		String[] args) {
	for (int i = 0; i < 10; i++) {
		System.out.println(
			"Hello World!");
	}
}
\end{lstlisting}
\vspace*{-1\baselineskip}
\end{footnotesize}
\end{minipage}
\end{lrbox}

\begin{figure}[t]
\centering
\vspace*{-0.8\baselineskip}
\subfloat[\label{fig:example:java}Java function.]{\usebox{\firstlisting}}
%
%
\subfloat[\label{fig:example:scratch}\Scratch code.]{\hspace{1ex}\includegraphics[width=0.23\columnwidth]{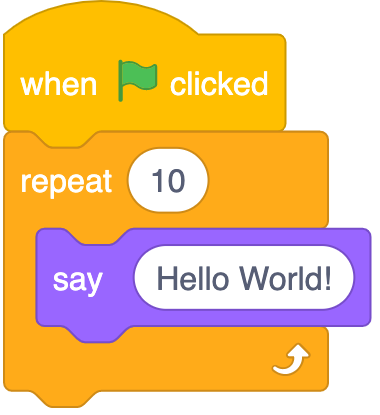}}\\
\subfloat[\label{fig:example:java:sequence}Java token sequence.]{
\parbox{1.0\columnwidth}{%
\footnotesize
\texttt{public} $\rightarrow$
\texttt{static} $\rightarrow$
\texttt{void} $\rightarrow$
\texttt{main} $\rightarrow$
\texttt{(} $\rightarrow$
\texttt{String} $\rightarrow$
\texttt{[} $\rightarrow$
\texttt{]} $\rightarrow$
\texttt{args} $\rightarrow$
\texttt{)} $\rightarrow$
\texttt{$\{$} $\rightarrow$ 
\texttt{for} $\rightarrow$
\texttt{(} $\rightarrow$
\texttt{int} $\rightarrow$
\texttt{i} $\rightarrow$
\texttt{=} $\rightarrow$
\texttt{0} $\rightarrow$
\texttt{;} $\rightarrow$
\texttt{i} $\rightarrow$
\texttt{$<$} $\rightarrow$
\texttt{10} $\rightarrow$
\texttt{;} $\rightarrow$
\texttt{i} $\rightarrow$
\texttt{++} $\rightarrow$
\texttt{)} $\rightarrow$
\texttt{$\{$} $\rightarrow$ 
\texttt{System} $\rightarrow$
\texttt{.} $\rightarrow$
\texttt{out} $\rightarrow$
\texttt{.} $\rightarrow$
\texttt{println} $\rightarrow$
\texttt{(} $\rightarrow$
\texttt{\"} $\rightarrow$
\texttt{Hello World!} $\rightarrow$
\texttt{\"} $\rightarrow$
\texttt{)} $\rightarrow$
\texttt{;} $\rightarrow$
\texttt{$\}$} $\rightarrow$
\texttt{$\}$}%
}
}\\
\subfloat[\label{fig:example:scratch:sequence}\Scratch token sequence.]{
\raisebox{-3mm}{\begin{scratch}\blockinit{When \greenflag clicked}\end{scratch} $\rightarrow$}
\begin{scratch}[scale=0.55]\blockrepeat{repeat \ovalnum{} times}{\blockspace[0.5]}\end{scratch}
\raisebox{-3mm}{
$\rightarrow$
\ovalnum{10} $\rightarrow$
\begin{scratch}\blocklook{say \ovalnum{}}\end{scratch} $\rightarrow$
\ovalnum{Hello World!}
}
}
\caption{\label{fig:example}Example code in text- and block-based format.}
\vspace*{-0.5\baselineskip}
\end{figure}

Just like for programs written in text-based programming languages,
there is a need to apply program analysis also to block-based
programs: Learners may benefit from automatically generated hints and
feedback, and programmers may benefit from code completion or bug
detection.
A popular approach to implement such analyses is to treat source code
like natural language, and thus benefit from the recent proliferation
of research on natural language processing (NLP) methods. At the core
of these methods lies the concept of language models, which capture
the probability distributions over sequences of words. As source code
has been observed to exhibit regularities that make it amenable to
natural language processing~\cite{Hindle2012}, the same models can
also be used to capture probability distributions for source
code. These models can, for example, predict common code sequences for
code completion~\cite{Raychev2014}, or identify unusual code sequences
for bug detection~\cite{Wang2016,ray2016naturalness}.

Language models are constructed by interpreting the source code as sequences of
\emph{tokens}, i.e., individual words or symbols separated by whitespace or
deliminator characters. The \java program in \cref{fig:example:java}
would thus be interpreted as the sequence of 39 tokens shown in
\cref{fig:example:java:sequence}. The \Scratch version of the same program
(\cref{fig:example:scratch}) results in a simple stream of only five tokens
(\cref{fig:example:scratch:sequence}). On the one hand, this difference can be
interpreted as strong reaffirmation of just how much block-based programming
reduces the cognitive overhead compared to text-based programming
languages~\cite{Bau2017}. On the other hand, it is unclear how this
simplification changes the resulting language models. Even when removing
\enquote{syntactic} tokens~\cite{Rahman2019}, the remaining tokens in the \java
example are intuitively at a lower level of abstraction than the tokens of the
\Scratch program, which contains less technical details such as modifiers or
types. Consequently, it is unclear how suitable language models are for program
analysis tasks on block-based programs.


In order to investigate whether block-based programs can be modelled
and analyzed using language models, we empirically investigate models
based on programs written in the \Scratch programming
language~\cite{maloney2010}, which is the most popular block-based
programming language and aims at young learners. There is a thriving
community of millions of users who share their programs, thus
providing large amounts of code, allowing us to perform an extrinsic
evaluation of the language models. In detail, the contributions of
this paper are as follows:
\begin{IEEEitemize}
	\item We describe and implement the process of creating \mbox{n-gram} models from
\Scratch programs. While there are various alternative neural models, n-gram
models have been shown to perform well for many tasks, and a sound
understanding requires interpretable models.
	\item We evaluate the suitability of n-gram models for the common tasks of
code completion, i.e., the prediction of which block will be used next in a
token stream, using a dataset of publicly shared \Scratch projects.
	\item We evaluate the ability of n-gram models to identify erroneous
solutions for \Scratch programming assignments.
        \item We investigate whether transformers, a popular deep learning model, can improve the performance of the completion task compared to n-gram models.
\end{IEEEitemize}

Our experiments confirm that block-based programs differ fundamentally
from text-based programs in a way that negatively affects their
predictability. However, there nevertheless are elements of syntax and
repetitiveness that make blocks sufficiently predictable to enable the
use of natural language models for block-based programming
languages. %

\section{Background}\label{sec:background}

Block-based programming languages have recently received increased attention
for teaching programming concepts to novices~\cite{mcgill2020} as well as for
industrial applications requiring end-user
programming~\cite{weintrop2017blockly,ritschel2020comparing,mayr2021considerations}. In this paper, we focus on the popular educational programming language
\Scratch~\cite{maloney2010}.

\subsection{The Scratch Programming Language}

\Scratch~\cite{maloney2010} is a block based programming language
 for young learners. \Scratch programs control the behavior of sprites
in an environment (stage); each sprite can implement multiple scripts.
\Cref{fig:example:scratch} exemplifies such a script: Scripts start with
event handlers (e.g., \begin{scratch}[scale=0.4]\blockinit{When \greenflag clicked}\end{scratch}) followed by
blocks that are executed after the event occurred. To support recognition
over recall blocks are color coded based on categories: control
structures are orange like the \includegraphics[height=1.1em]{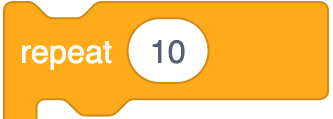} block in
\cref{fig:example:scratch}, blocks affecting the visual appearance of sprites are purple (e.g., \begin{scratch}[scale=0.4]\blocklook{say \ovalnum{Hello World!}}\end{scratch}), etc.
Blocks are further divided into different shapes, such as stackable blocks (statements) and round or diamond-shaped reporter blocks that fit into holes in other blocks (expressions).
%
Blocks may have free text spaces for numbers and strings like in \begin{scratch}[scale=0.4]\blocklook{say \ovalnum{Hello World!}}\end{scratch}, and drop-down menus to select pre-set options. 
\Scratch enables a remix culture~\cite{Bau2017} where users share their programs, and others clone and enhance them.


Even though the block shapes prevent syntactical errors, building
programs can nevertheless be challenging: learners may struggle to
implement functionality or may have
misconceptions~\cite{hermans2016a,hermans2016b,techapalokul2017a,commonBugs},
and programmers may miss the convenience and support of modern
programming environments.
%
%
As a consequence, various analysis tools have been proposed, mainly
implementing traditional program analyses such as linting
(e.g.,~\cite{Litterbox,techapaloku2017b}) or automated testing
(e.g.,~\cite{johnson2016itch,stahlbauer2019testing}). However, analysis tools
using NLP methods are, to the best of our knowledge, not available
yet.


\subsection{N-gram Models}\label{sec:ngram}

Probabilistic language models are used to assign probabilities to sequences of tokens in a given language.
N-gram language models are based on the Markov assumption, which states that the probability of a sentence $s$ can be estimated based on a chain of probabilities for all its tokens $w_1 \ldots w_n$ to occur.
N-gram models further simplify this idea and assume that each word actually depends only on its $n-1$ preceding words, that is, on its \emph{local context}.
Given $n=3$ and $s = \langle w_1 w_2 w_3 w_4 \rangle$,
an n-gram model thus estimates the probability of $s$ as follows:
\begin{equation*}
P (s) \approx P(w_1) \times P(w_2 | w_1) \times P(w_3 | w_1 w_2) \times P(w_4 | w_2 w_3)
\end{equation*}
In contrast to regular Markov chains, the probability of $w_4$ is estimated considering only the context $w_2 w_3$, not $w_1 w_2 w_3$.
The factors of the product are conditional probabilities estimated using the count of occurrences of tokens in the training data.
%
%
For example, the probability $P(c | ab)$ with local context $ab$ has a probability of $1$, if only $c$ follows $ab$ in the training data.

A probability of $1$ is rather unrealistic in practice:
It is more likely that the training data did not contain all possible tokens that may follow the local context.
%
Therefore, 
smoothing algorithms shift the raw probabilities based only on the counts of
the n-grams, so that other n-grams that are not part of the training data are
also assigned probabilities $> 0$.
This way, the model is not \enquote{infinitely surprised} by n-grams other than those present in the training data set.
A popular smoothing mechanism is modified Kneser-Ney smoothing~\cite{Chen1999}, which has been reported to perform best in natural language contexts~\cite{Chelba2014} and was also used in prior work on programming languages~\cite{Hindle2012}.

\subsection{Deep Learning Models: Transformer}

While n-gram models are valued for their simplicity and ease of interpretation, research has recently shifted towards neural approaches.
Vaswani et al.~\cite{vaswani2017attention} proposed the transformer architecture to enable capturing long range information during automated natural language translation.
The transformer design makes use of the encoder-decoder architecture~\cite{vaswani2017attention}:
In the encoding part the model learns weight matrices for different word relations that encode how strong a word-encoding is influenced by every other word within the sequence~\cite{vaswani2017attention}.
During decoding the next generated token is influenced not only by the previously generated output, but also by the weight matrices over the whole input sequence~\cite{vaswani2017attention}.
%
Transformers allow for self-supervised learning, e.g., by masking random tokens in input sequences and training the transformer to predict the missing words~\cite{devlin2019bert}.
These models tend to require a more computationally expensive training process compared to \mbox{n-gram} models, yet it has been shown that a transformer trained on a large dataset (e.g., BERT~\cite{devlin2019bert}, CodeBERT~\cite{feng2020codebert}) can be used without or with only little fine-tuning to assist with tasks in the domain of source code processing~\cite{ciniselli2021,Chirkova2021,degiovanni2022mubert,Svyatkovskiy2020}.

\subsection{Program Analysis with Language Models}\label{sec:programanalysis}

Hindle et al.\ introduced the \enquote{naturalness hypothesis} based on which they proposed to use n-gram models to model source code~\cite{Hindle2012}.
%
As an initial application, they presented a simple code completion based on the
n-gram model, which suggests a ranking of the most likely tokens based on the
local context of the completion.
That is, one maximizes the probability of the complete sequence by choosing the most probable n-gram based on the given local context.

The idea of benefiting from concepts of NLP and combining it with other techniques has been taken up successfully in other work, for example, in the area of code completion~\cite{Nguyen2013,Raychev2014}.
However, further investigations of the naturalness hypothesis have also shown that large parts of the naturalness of code are due to syntactic elements such as parentheses or semicolons~\cite{Rahman2019}.
Even though source code is less natural than previously thought, regularities can still be found in the source code even after removing certain syntactic elements.
%
%
In particular, despite the limitations of the naturalness hypothesis, n-gram
models have been determined to be capable of representing source code very
well, often better than deep neural networks~\cite{hellendoorn2017deep}, when
appropriately configured.

If source code is regular, then irregularities in the source code are
suspicious:
Ray et al.\ demonstrated 
that buggy code has a higher entropy than correct source code~\cite{ray2016naturalness}.
%
This insight enables the application of n-gram models to identify bugs in
source code. In particular,
Wang et al.\ introduced \bugram, an
automated approach for finding bugs based on \mbox{n-gram} models~\cite{Wang2016}.
Given a specific project, \bugram trains an \mbox{n-gram} model on the source code, calculates probabilities for all sequences in the source code, and
reports sequences with low probability as potential bugs.
%
%


Due to their specific structure, transformers open up further possibilities for code analysis.
They can be used to jointly learn from code and natural language by training the model on sequences containing both tokenized code and its documentation to enable code search using natural language descriptions or generating documentation~\cite{feng2020codebert}.
Those pre-trained models can then also be applied to code-only tasks like identifying buggy code~\cite{pan2021} and generating potential fixes for it~\cite{mashhadi2021}.


\section{Language Models for Scratch}\label{sec:approach}

\Scratch differs from text-based programming languages, to which
language models have been previously applied. Thus, we first need to define
how to tokenize \Scratch programs.
We then describe how n-gram models are generated and how they are applied for code completion
and bug finding.
We further describe how we obtain the transformer, and how it can be used for code completion.

\subsection{Tokenizing \Scratch Programs}\label{sec:model}


Tokenizing text-based programming languages is straightforward, e.g., by directly lexing
the source code. It is less obvious, however, how to tokenize \Scratch
programs: a \Scratch program consists of a ZIP-file containing resources
(images, sounds) as well as a text file in JavaScript Object Notation (JSON)
format describing the code. The JSON file describes a program in terms of the
\emph{targets} (i.e., stage and sprites), and each target consists of its name,
variables, lists, messages, sounds, costumes, scripts, procedures (i.e., custom
blocks), and blocks. The blocks are listed in an arbitrary order (e.g., the
order in which they were inserted in the program), and each block consists of a
unique identifier as well as the identifiers of the parent and successor
blocks, as well as any parameter blocks. The block identifiers and their parent/child
relations are used in the \Scratch virtual machine to create a syntax-tree-like
representation. Although the JSON format is specific to \Scratch, other
block-based languages represent programs similarly; for example,
\textsc{Snap!} encodes blocks in XML~\cite{harvey2013snap}.
In order to tokenize a program, we first use the parser
provided by the \litterbox~\cite{Litterbox} analysis framework and create the
abstract syntax tree for that program. We then traverse the syntax tree in
preorder, adding each traversed node that represents a concrete block to the
token stream. The resulting sequence of tokens is illustrated in
\cref{fig:example:java:sequence}.

To reduce the vocabulary size and avoid out-of-vocabulary issues~\cite{Karampatsis2020}, we treat literals as follows~\cite{babii2019modeling}: First, we do not include string and number literals. On one hand, predicting literals is very difficult; on the other hand, text and numbers entered by users are usually very dependent on the use case. Second, similar to prior work~\cite{ahmed2018compilation,xu2019commit}, we generalize the occurrence of concrete variables to the occurrence of the generic variable block \setscratch{scale=0.7}\ovalvariable{var} and calls to self defined blocks as procedures call \begin{scratch}[scale=0.4]\blockmoreblocks{call}\end{scratch}.
Since most programs actually define only a few variables and procedures, this
is on the one hand potentially not a very large loss of information, but on the
other hand simplifies generalization across project boundaries.

Even though \Scratch treats the drop-down menus that some blocks include as
individual blocks in its internal representation, we do not include these as
tokens as they are inseparable and are tailored to
the specific block and use case.
Thus, overall we only include statement and expression blocks
as tokens, which results in a vocabulary size of 137 blocks.  As usual
in NLP, we introduce structural blocks for the beginning and ending of
scripts or sprites. Consequently, the remaining token sequence of the
script in \cref{fig:example:scratch} 
looks as follows:

\begin{center}
\centering
\raisebox{-3mm}{
  \textsf{\footnotesize Begin Script} $\rightarrow$
  \begin{scratch}[scale=0.45]\blockinit{When \greenflag clicked}\end{scratch} $\rightarrow$
}
\hspace*{-2mm}
\begin{scratch}[scale=0.45]\blockrepeat{repeat \ovalnum{} times}{\blockspace[0.5]}\end{scratch}
\hspace*{-2mm}
\raisebox{-3mm}{
  $\rightarrow$
  \begin{scratch}[scale=0.45]\blocklook{say \ovalnum{}}\end{scratch} $\rightarrow$
  \textsf{\footnotesize End Script}
}
\end{center}

%

%

\subsection{Code Completion using N-gram Models}\label{sec:approach-completion}

 \begin{figure}[tb]
	 	\centering
	 	\subfloat[\label{fig:example_code:compl}Incomplete script during development.]{
	 	\includegraphics[scale=0.22]{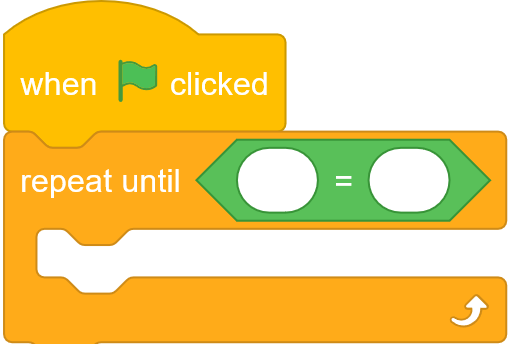}
	    }
	    \hspace*{2pt}
	    \subfloat[\label{fig:example_code:bug}Completed script using an uncommon condition in the repeat block.]{
	 	\includegraphics[scale=0.22]{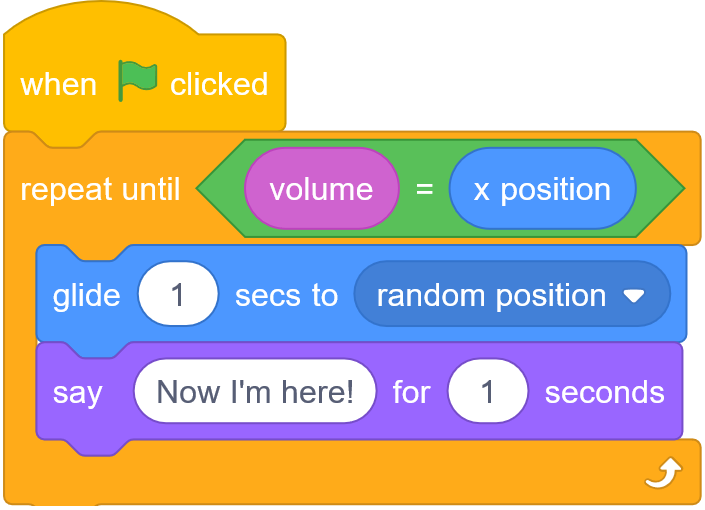}
	 	}
	 	\caption{Example of a \Scratch script.}
\end{figure}

N-gram models are the most fundamental type of language models.  They
work with relatively small amounts of training data compared to more
modern deep learning approaches, and have been successfully applied to
various software engineering tasks (cf.~\cref{sec:programanalysis}),
and are therefore implemented as a baseline for the use of language
models in our work.
The first task on which we apply the n-gram model is code completion.
The idea is to provide code completion for the next block when interpreting the preceding code linearly (i.e., in the order of the token sequence).
\cref{fig:example_code:compl} shows a simple \Scratch program during development.
The existing code tokens would be interpreted in the following order:

\begin{center}
\setscratch{scale=0.43}
\raisebox{-3mm}{%
  \textsf{\footnotesize Begin Script} $\rightarrow$
  \begin{scratch}\blockinit{When \greenflag clicked}\end{scratch} $\rightarrow$
}\hspace*{-1mm}
\begin{scratch}\blockrepeat{repeat until \boolempty[1em]}{\blockspace[0.5]}\end{scratch}
\raisebox{-3mm}{%
  \(\rightarrow\)
  \booloperator{\ovalnum{} = \ovalnum{}} $\rightarrow$
  \textsf{\footnotesize End Script}
}
\vspace*{0.1cm}
\setscratch{scale=0.5}
\end{center}

For a 3-gram model, \includegraphics[height=1.1em]{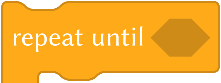} and
\setscratch{scale=0.6}\booloperator{\ovalnum{} = \ovalnum{}} represent the
local context to be completed by the model.
Thus, in this case the code completion shall suggest blocks that are usually
used for equality comparisons in while-loops, such as \setscratch{scale=0.7}\ovalvariable{var}\setscratch{scale=0.6}, \ovallook{costume \selectmenu{number}}, or \ovaloperator{round \ovalnum{}}\setscratch{scale=0.55}. 
To this end, we build a \enquote{general} n-gram model using modified
Kneser-Ney smoothing on a large set of \Scratch programs as described
in \cref{sec:model}.

A special case for code completion in \Scratch exists in the case of procedure
definition blocks:
Analogous to method headers in \java, these blocks build the header of a newly created custom block.
%
%
%
Since procedure definitions are not added using drag and drop like
other blocks, but using a dedicated dialogue to set the name and
select possible parameters, we exclude procedure definitions from
predictions. 
%
Another special case is the behavior of code completion when the model predicts
the end of a block sequence based on the \textsf{\footnotesize End Script} blocks that were
observed during training. In this case, instead of this prediction, which has
no value for a user, the completion returns a prediction for a new first block,
i.e., the completion for the context at the beginning of a script.

To the best of our knowledge, this is the first implementation of code
completion for block-based programming languages. Therefore, we
implemented a simple code completion based purely on the probabilities
of the n-gram model, which allows us to evaluate how well the language
model itself represents the language, and provides a baseline for
further research.

\subsection{Bug Finding using N-gram Models}\label{sec:approach-bugram}


Language models can reflect that buggy code is less regular than non-buggy
code~\cite{ray2016naturalness}. For example, \cref{fig:example_code:bug} shows a \Scratch script including a
very unlikely condition in the loop block \includegraphics[height=1.1em]{img/repeatuntil}:
Comparing the audio volume \ovalsound{volume} to the \mbox{x position} of a sprite
\ovalmove{x position} is very unlikely to be a meaningful comparison.
Ideally, a bug finding approach would be capable of capturing these and other
irregularities in the source code.

This principle has been applied in the \bugram
approach~\cite{Wang2016} (cf. \cref{sec:programanalysis}).
%
%
As \Scratch projects are much smaller than \java or
\python projects, and likely contain less repetitive API
usages, \bugram is not directly applicable. However, a common scenario
in an educational context is that students implement a task for which
there are one or more model solutions.
Thus, we train an n-gram model on model solutions or known good
student solutions, and then assess the probabilities of all sequences
in the student solutions, reporting those with a particularly low
probability.
Since this model is not trained on the entire code base (i.e., all
student solutions), unlike in the \bugram approach, the use of a
smoothing algorithm is necessary, for which we again use modified
Kneser-Ney smoothing.
Additionally, we skip any preprocessing applied by \bugram, such as including
the whole path in method names or skipping tokens with a particularly low count
as proposed by
Wang et al.~\cite{Wang2016} when parsing the source code, due to the
simple structure of \Scratch.
Thus, we structurally use the same model for code completion and the bug finding task.

%

%
%



\subsection{Code Completion using a Transformer}\label{sec:transformer}

While n-gram models are simple, light weight, and easily
interpretable, transformers are more contemporary and often yield
better results~\cite{ciniselli2021,kim2021}. However, they are limited to tasks
where very large amounts of training data are available, which in our study
includes only the code completion task.



Using the script tokenization (\cref{sec:model})
we generate one token sequence \([t_0, \dots, t_n]\) per sprite by
concatenating the token sequences from all procedure definitions
followed by scripts.
Since users can place their blocks freely on the canvas, we tokenize the
scripts in the order which they appear in the project file.
This usually represents the order in which the user created them.
For training we used the RoBERTa~\cite{liu2019roberta}
implementation provided in the \python \emph{transformers}
library~\cite{huggingfaceTransformers2020}.  As identifiers are
removed from the token sequence, the model does not have to handle
out-of-vocabulary scenarios and the vocabulary is reduced to 137
different tokens.  Therefore, a simple word-level tokenizer that
assigns a numeric identifier for each token is used.



The model is trained using a masked language
model~\cite{devlin2019bert} with the default RoBERTa approach of randomly
masking tokens~\cite{liu2019roberta}. In
this approach tokens are randomly replaced by a placeholder
\texttt{[MASK]} for which the original token then has to be predicted
based on the surrounding context:

\begin{center}
\centering
\textsf{\footnotesize Begin} $\rightarrow$
\begin{scratch}[scale=0.5]\blockinit{When \greenflag clicked}\end{scratch} $\rightarrow$
\texttt{[MASK]} $\rightarrow$
\begin{scratch}[scale=0.5]\blocklook{say \ovalnum{}}\end{scratch} $\rightarrow$
\textsf{\footnotesize End}
\end{center}

When using the trained model for code completion, a sequence of up to
the last \(m - 1\) tokens of the existing code is extracted and the
additional \texttt{[MASK]} token appended.  The model then predicts a
probability for each token to replace this mask, so that a top-$x$
selection of suggestions can be presented to the user.
Analogous to the completion with \mbox{n-gram} models, procedure definitions are excluded as described in \cref{sec:approach-completion}, and predictions for the end of a script are replaced by suggestions for a new script.

\section{Evaluation}\label{sec:evaluation}

As a baseline to gain an understanding of the applicability of language
models in the context of block-based programming languages, we
experimentally examine n-gram models on \Scratch programs from two
opposite angles: First, we consider how well n-gram models capture the
regularities of \Scratch programs by looking at the highest
probabilities encoded in the model, using the task of code
completion. Second, we consider how well n-gram models detect
deviations from common patterns by looking at the lowest probabilities
encoded in the model, using the task of bug finding. Finally, we
investigate whether predictions can be improved using deep learning
models.  This leads to the following research questions:
\begin{itemize}
	\item \textbf{RQ1, Completion:} How well does code completion based on n-gram models perform on \Scratch source code?
	\item \textbf{RQ2, Bug Finding:} How well does bug finding based on n-gram models perform on \Scratch source code?
	\item \textbf{RQ3, Model Comparison:} Do transformer-based deep learning models improve over n-gram models?
\end{itemize}

%

\subsection{RQ1: Code Completion with N-gram Models}\label{sec:rq1}

\subsubsection{Experimental Setup}\label{subsec:setup_compl}

To create a general model for the purpose of code completion, we
trained an n-gram model on \num{100000} randomly selected \Scratch
programs.
Between May 2021 and February 2022 we retrieved the \num{10000} most recently
publicly shared \Scratch programs each day using the REST API of the \Scratch
website, resulting in a total 2.7 million projects. From these, we filtered
projects with less than 10 blocks, as these very often represent projects in
which children focused on arts and drawing, e.g., drawing a background and
arranging sprites on it, rather than coding.
Furthermore, we excluded remixes (i.e., copied and modified programs).
From the resulting 1.1 million projects, we then randomly sampled
\num{110000}, which we split into a training set of \num{100000}
projects, and an evaluation set of \num{10000} projects.
%
Comparing the number of blocks between projects in the training and
evaluation sets shows that there is no significant difference
($p = \pvalueBlocks$ using a Mann-Whitney U test~\cite{mann1947test}),
thus confirming that the two datasets are drawn from the same overall
distribution.

To choose a suitable value for the sequence length $n$, we trained the model for $n = \{1, 2, 3,
4\}$.
We used $4$ as the upper bound for two reasons:
First, we observed only marginal improvements for larger $n$, which is in line
with findings in prior work~\cite{hellendoorn2017deep,Hindle2012}.
Second, the models require substantially more memory and computation time as $n$
increases. This is particularly relevant in the bug-finding use case, where a
custom model is trained, e.g., for a given set of programs, in order to identify
unusual sequences. In the context of finding bugs in
programming assignments, which tend to use relatively small and simple \Scratch
programs, training a complex model may not be worthwhile.
%
We use the same \num{100000} randomly selected \Scratch programs for each
n-gram model.



The projects from the evaluation data set are broken down into sets of local contexts,
each consisting of $n-1$ blocks.
Every context is given to the completion engine that is asked to predict the
next block for this context.
The completion engine returns the top $x = \{1, 2, 3, 5, 10\}$ blocks
ranked by their probability, and
we evaluate the code-completion suggestions in terms of top-$x$ accuracy, i.e.,
the ratio of suggestions that contained the actual block in the original program.
We consider top-$x$ accuracy for varying $x$ and $n$, and we evaluate the
influence of block frequency, category, and shape on top-$x$ accuracy.

\subsubsection{Threats to Validity}\label{subsec:threats_completion}

Threats to external validity arise as
results may not generalize to projects outside our dataset. We
confirmed that in terms of size the sample is a valid representation
of publicly shared projects; however, unfinished, unshared programs
might have other properties. To avoid skewing results with very
similar code we used only original projects and excluded
remixes. Threats to internal validity may arise from our
implementation: Although we tested and validated all code thoroughly,
our implementation may confound the studied measurements and
relationships. For example, rare aspects of the \Scratch program
representation not encountered during testing may be misrepresented.
Threats to construct validity may arise from our choice of top-$x$ accuracy as
metric rather than precision or recall. This choice is based on the use case:
We assume that each suggestion in the top-$x$ is equally useful.
Indeed a deployed code-completion engine can suppress low-confidence
predictions, and a user does not have to accept incorrect suggestions.

\subsubsection{Results}\label{subsec:results_compl}

\begin{table}[]
\centering%
\caption{Accuracy of code completion in top x suggestions for different values of n.}\label{tab:completion-accuracy}
\begin{tabular}{lrrrrr}
\toprule
            & Top 1           & Top 2           & Top 3          & Top 5         & Top 10  \\
\midrule
1-gram      & \accOneOne      & \accOneTwo      & \accOneThree   & \accOneFive   & \accOneTen   \\
2-gram      & \accTwoOne      & \accTwoTwo      & \accTwoThree   & \accTwoFive   & \accTwoTen   \\
3-gram      & \accThreeOne    & \accThreeTwo    & \accThreeThree & \accThreeFive & \accThreeTen \\
4-gram      & \accFourOne     & \accFourTwo     & \accFourThree  & \accFourFive  & \accFourTen  \\
transformer & \accTransOne    & \accTransTwo    & \accTransThree & \accTransFive & \accTransTen \\
\bottomrule
\end{tabular}

\end{table}

\Cref{tab:completion-accuracy} shows the top-$x$ accuracy of code completion for $n =
\{1, 2, 3, 4\}$.
Top-$x$ accuracy is defined as the sum of all true positive predictions
per block divided by the total number of predictions. A prediction is considered
a true positive if the set of top-$x$ suggestions contains the actual block to
be predicted.
By definition, top-$x$ accuracy monotonically increases with increasing $x$. We
also observe that it increases with increasing $n$.
Since 4-grams perform best, we use 4-grams for the rest of RQ1. Furthermore, we use top-3 accuracy for subsequent
evaluations as 3 is a reasonable number for suggestions in the
\Scratch user interface (e.g., in the \enquote{backpack} of code snippets) satisfying the design philosophy of \Scratch to keep the cognitive load low~\cite{Bau2017}.




The overall best predicted block is \begin{scratch}[scale=0.4]\blockinit{When
\greenflag clicked}\end{scratch} with a top-3 accuracy of \textpercent{95.52}. With an
occurrence rate of \textpercent{6.03} this is the second most frequent block overall
in the
evaluation data. Consequently, it is likely that occurrence has an influence on
the performance of the prediction.
We note that the top blocks in other categories
show a substantially lower top-3 accuracy,
thus there appears to be an influence also of the category.
Finally,
we observe that generally oval and diamond shaped blocks all have
particularly high accuracy values, suggesting that the shape of
blocks also contributes to the prediction.  In order to better
understand what determines the overall prediction performance, we
therefore investigate the influence of these three aspects: (1) the
frequency with which blocks occur in practice; (2) the category the
blocks belong to (e.g., motion, looks, \textellipsis); and (3) the shape of the
blocks (e.g., regular stackable blocks, event handler blocks, \textellipsis).

 \begin{figure*}[tb]
   \centering
   \subfloat[\label{fig:scatter-recall}Top-3 accuracy vs.\ occurrence of blocks, grouped by category and shape of blocks. (Note the differently scaled x-axes.)]{
     \includegraphics[width=0.66\linewidth]{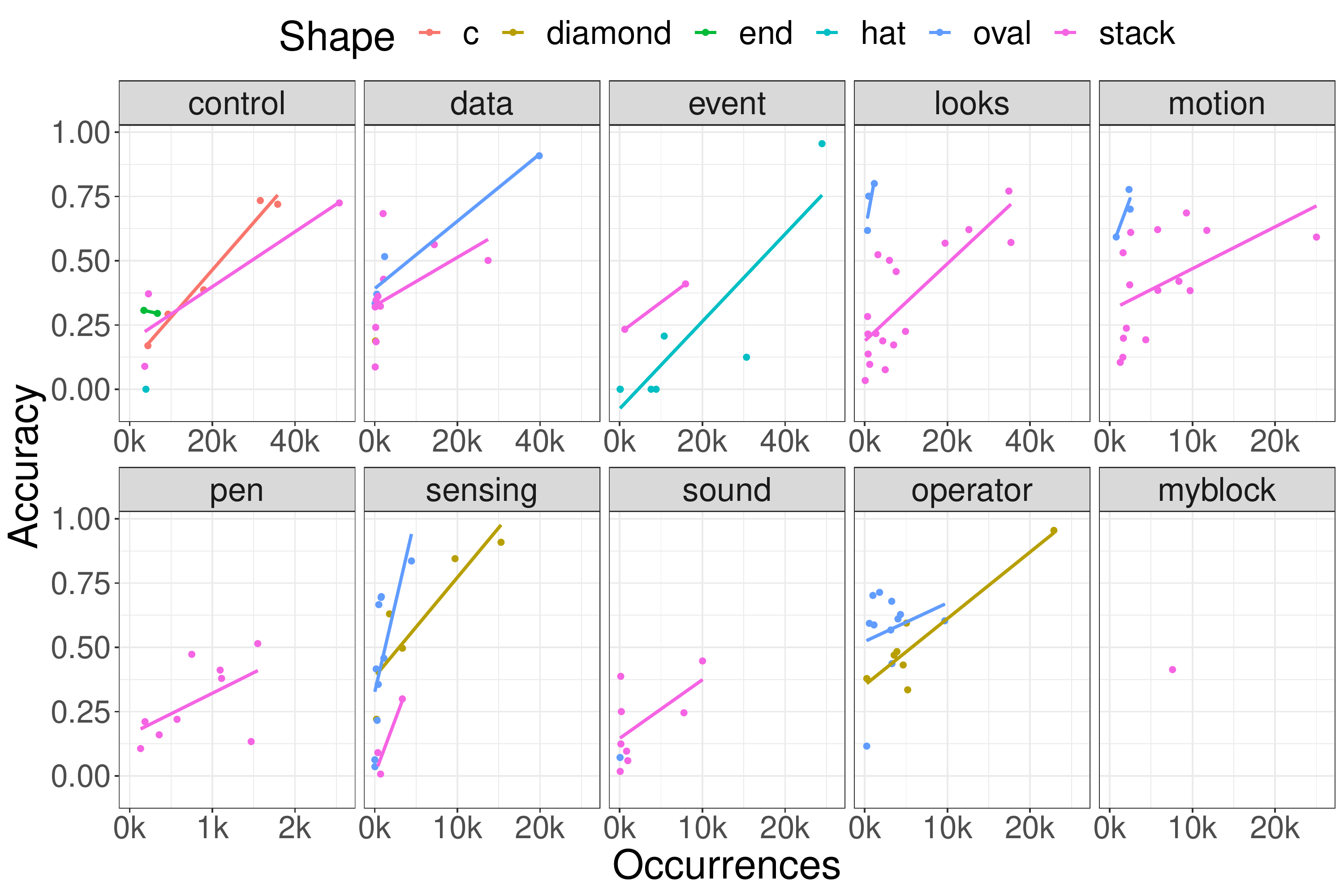}
   }
   \hspace*{2pt}
   \subfloat[\label{tab:regressionmodel}General linear model.]{
\resizebox{.27\textwidth}{!}{%
\sisetup{table-format=1.3}
\begin{tabular}[b]{lrS[table-auto-round]}
\toprule
Variable & \multicolumn{1}{c}{Coefficient} & \multicolumn{1}{c}{$p$} \\
\midrule
\textbf{Intercept}   & \num{ 2.942e-01} & {$<$\,0.001} \\
\midrule
\textbf{Occurrences} & \num{ 1.488e-05} & {$<$\,0.001} \\
\midrule
\textit{Category}\\
control              & \num{-1.050e-01} & 0.21638 \\
data                 & \num{ 9.807e-03} & 0.83368 \\
event                & \num{-4.618e-02} & 0.63354 \\
looks                & \num{ 5.690e-03} & 0.89679 \\
\textbf{motion}      & \num{ 1.078e-01} & 0.01852 \\
pen                  & \num{ 5.715e-02} & 0.34205 \\
sensing              & \num{-2.677e-02} & 0.62158 \\
sound                & \num{-9.485e-02} & 0.10814 \\
operator             & \num{ 1.275e-02} & 0.83089 \\
myblock              & \num{ 7.960e-02} & 0.61187 \\
\midrule
\textit{Shape}\\
c                    & \num{-2.339e-02} & 0.79664 \\
\textbf{diamond}     & \num{ 1.549e-01} & 0.03448 \\
end                  & \num{ 3.617e-02} & 0.76269 \\
\textbf{hat}         & \num{-2.862e-01} & 0.00303 \\
\textbf{oval}        & \num{ 1.921e-01} & 0.00157 \\
stack                & \num{-7.354e-02} & 0.14629 \\
\bottomrule
\end{tabular}%
}

%
%
%
%
%

   }
   \caption{Influence of block occurrence, category, and shape on code-completion accuracy.}
 \end{figure*}

\Cref{fig:scatter-recall} summarizes these three aspects and their influence on
the top-3 accuracy for $n = 4$: The plot is split into facets based on the 10
different block categories in which the blocks are sorted in the \Scratch user
interface. For each category, blocks are plotted based on the number of
occurrences in the training data (x-axis) and the resulting top-3 accuracy
(y-axis). Data points are color-coded based on their shape.
In particular, \emph{hat
blocks} represent event handlers, \emph{stack blocks} are regular statements,
\emph{oval} blocks represent reporters returning numerical or textual data,
\emph{diamond} shaped blocks represent Boolean values, \emph{c}-shaped blocks
are control structures such as loops and if-conditions, and \emph{stop} blocks
are statement blocks that cannot have successors.
The \enquote{myblocks} category only contains one block because we generalize
identifiers (see~\cref{sec:approach-completion}), and only predict calls to
these self-created blocks, not their creation. Consequently, data points for
procedure definition blocks and their possible parameters are not included.
%
%
To better understand the influence of block frequency, category, and shape on
accuracy, we used multiple linear regression to model this relationship.
We include occurrence as a continuous variable and category and shape as
categorical variables. Since we are interested in whether the accuracy for
particular categories and/or shapes, independently of occurrence, differs
significantly from the average accuracy, we used deviation coding---comparing
each level to the grand mean.
\Cref{tab:regressionmodel} shows the results of the regression analysis.
%
%

\begin{table}[]
\centering%
\caption{Completion accuracy by category for n=4, x=3.}\label{tab:completion-acc-by-group}
\begin{tabular}{lrrr}
\toprule
Group & Occurrences & Accuracy & Acc. Transformer \\
\midrule
sound & \printpercent{0.024832701666393637} \DrawPercentageBar{0.024832701666393637} & \printpercent{0.32934874262189373} \DrawPercentageBar{0.32934874262189373} & \printpercent{\accTransSound} \DrawPercentageBar{\accTransSound} \\
pen & \printpercent{0.008894248238332843} \DrawPercentageBar{0.008894248238332843} & \printpercent{0.33998061210358677} \DrawPercentageBar{0.33998061210358677} & \printpercent{\accTransPen} \DrawPercentageBar{\accTransPen} \\
myblocks & \printpercent{0.009353679701135523} \DrawPercentageBar{0.009353679701135523} & \printpercent{0.41322096391888335} \DrawPercentageBar{0.41322096391888335} & \printpercent{\accTransMyblocks} \DrawPercentageBar{\accTransMyblocks} \\
event & \printpercent{0.15295618896058866} \DrawPercentageBar{0.15295618896058866} & \printpercent{0.479944596999541} \DrawPercentageBar{0.479944596999541} & \printpercent{\accTransEvent} \DrawPercentageBar{\accTransEvent} \\
motion & \printpercent{0.12166188554121149} \DrawPercentageBar{0.12166188554121149} & \printpercent{0.5149330795553486} \DrawPercentageBar{0.5149330795553486} & \printpercent{\accTransMotion} \DrawPercentageBar{\accTransMotion} \\
looks & \printpercent{0.20740805520075184} \DrawPercentageBar{0.20740805520075184} & \printpercent{0.5389544447677699} \DrawPercentageBar{0.5389544447677699} & \printpercent{\accTransLooks} \DrawPercentageBar{\accTransLooks} \\
control & \printpercent{0.21178312371516234} \DrawPercentageBar{0.21178312371516234} & \printpercent{0.5880447362758155} \DrawPercentageBar{0.5880447362758155} & \printpercent{\accTransControl} \DrawPercentageBar{\accTransControl} \\
operator & \printpercent{0.09596574833748628} \DrawPercentageBar{0.09596574833748628} & \printpercent{0.663209775130917} \DrawPercentageBar{0.663209775130917} & \printpercent{\accTransOperator} \DrawPercentageBar{\accTransOperator} \\
data & \printpercent{0.11372345182066654} \DrawPercentageBar{0.11372345182066654} & \printpercent{0.6800896792990285} \DrawPercentageBar{0.6800896792990285} & \printpercent{\accTransData} \DrawPercentageBar{\accTransData} \\
sensing & \printpercent{0.05342091681827084} \DrawPercentageBar{0.05342091681827084} & \printpercent{0.7359525950519933} \DrawPercentageBar{0.7359525950519933} & \printpercent{\accTransSensing} \DrawPercentageBar{\accTransSensing} \\
\bottomrule
\end{tabular}
\end{table}

\begin{table}[]
\centering%
\caption{Completion accuracy by shape for n=4, x=3.}\label{tab:completion-acc-by-shape}
\begin{tabular}{lrrr}
\toprule
Shape & Occurrences & Accuracy & Acc. Transformer \\
\midrule
end & \printpercent{0.012504189420972735} \DrawPercentageBar{0.012504189420972735} & \printpercent{0.29897516752069375} \DrawPercentageBar{0.29897516752069375} & \printpercent{\accTransShapeEnd} \DrawPercentageBar{\accTransShapeEnd} \\
hat & \printpercent{0.13672421781046074} \DrawPercentageBar{0.13672421781046074} & \printpercent{0.4757437297789313} \DrawPercentageBar{0.4757437297789313} & \printpercent{\accTransShapeHat} \DrawPercentageBar{\accTransShapeHat} \\
stack & \printpercent{0.520062397728841} \DrawPercentageBar{0.520062397728841} & \printpercent{0.5177934256727621} \DrawPercentageBar{0.5177934256727621} & \printpercent{\accTransShapeStack} \DrawPercentageBar{\accTransShapeStack} \\
c & \printpercent{0.12194911578573822} \DrawPercentageBar{0.12194911578573822} & \printpercent{0.5994038597554815} \DrawPercentageBar{0.5994038597554815} & \printpercent{\accTransShapeC} \DrawPercentageBar{\accTransShapeC} \\
oval & \printpercent{0.11507723320781499} \DrawPercentageBar{0.11507723320781499} & \printpercent{0.7492959857805189} \DrawPercentageBar{0.7492959857805189} & \printpercent{\accTransShapeOval} \DrawPercentageBar{\accTransShapeOval} \\
diamond & \printpercent{0.09368284604617234} \DrawPercentageBar{0.09368284604617234} & \printpercent{0.7523609101670393} \DrawPercentageBar{0.7523609101670393} & \printpercent{\accTransShapeDiamond} \DrawPercentageBar{\accTransShapeDiamond} \\
\bottomrule
\end{tabular}
\end{table}

\Cref{fig:scatter-recall} suggests differences between categories, which are
summarized in \cref{tab:completion-acc-by-group}: Blocks of the sensing, data,
operator, and control categories are predicted with a substantially higher
accuracy than, for example, blocks from the sound or pen categories. One of the
reasons for this lies in differences in the frequency of occurrence. For
example, blocks of the pen or sound categories appear much less frequently
than, for example, blocks from the motion or looks categories.
The importance of occurrence can also be observed within categories, not just
across categories. The fitted lines in \cref{fig:scatter-recall} very clearly
demonstrate that across all categories and shapes, the number of occurrences
has a positive influence on the accuracy: The more frequently a block occurs in
practice, the higher its probability of being predicted correctly. This is, for
example, confirmed by the high occurrence and accuracy of the best predicted
block of the event category (cf.~\cref{tab:completion-acc-by-group}), i.e.,
\begin{scratch}[scale=0.4]\blockinit{When \greenflag clicked}\end{scratch}.
\Cref{tab:regressionmodel} confirms that the occurrence has a significant
effect on the accuracy of the prediction, although the small coefficient of the
regression model indicates the influence is small.

However, not all differences between the categories can be explained through
the numbers of occurrences. 
For example, the
categories sensing and operators can be predicted relatively well
(\cref{tab:completion-acc-by-group}), even though blocks in these categories
occur less frequently overall compared to, e.g., motion blocks.
\Cref{tab:regressionmodel} confirms some influence of the category; in
particular, the motion category has a significant influence on the prediction
accuracy.
However, we observe 
that the
well-predictable sensing and operator categories differ from others in an
important property: they all contain a particularly large proportion of
diamond-shaped and oval-shaped blocks, which represent expressions rather than
statements (\textpercent{100} expression blocks in operator, \textpercent{83} in sensing). Pen and
sound blocks, for example, consist of almost only regular stack blocks.
%
\Cref{fig:scatter-recall} generally demonstrates with the different colors and
fitted lines that stack blocks appear to be more difficult to predict than, for
example, oval blocks (cf.\ categories data, looks, motion, sensing) or
diamond-shaped blocks (cf.\ sensing).
\cref{tab:completion-acc-by-shape} supports this impression using
the top-3 accuracy values grouped by the shape of the blocks.
\Cref{tab:regressionmodel} confirms that oval and diamond shapes have a
significant positive influence on the prediction, whereas hat blocks have a
significant negative influence.

\begin{table}
	\centering
	\caption{Top 3 predictions for the example in \cref{fig:example_code:compl}.}\label{tab:recommended}
	\begin{tabular}{lr}
		\toprule
		Block & Confidence\\
		\midrule
		\setscratch{scale=0.7}\ovalvariable{var}& \SI[round-mode=places, round-precision=2]{80.31181608409956}{\percent} \\
		\ovallook{costume \selectmenu{number}} & \SI[round-mode=places, round-precision=2]{5.665420458992064}{\percent} \\
		\setscratch{scale=0.65}\ovalsensing{answer} & \SI[round-mode=places, round-precision=2]{5.439873290190319}{\percent}   \\
		\bottomrule 
	\end{tabular}
\end{table}

\begin{figure}
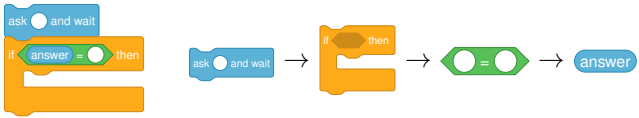
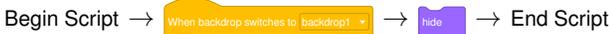

\vspace*{-8pt}

\begin{center}
\centering
\subfloat[\label{fig:idiom}Code for processing user input (left) and corresponding 4-gram (right). The probability of this 4-gram is \textpercent{97.56}.]{
  \begin{minipage}{0.25\columnwidth}
    \begin{scratch}[scale=0.45]
	  \blocksensing{ask \ovalnum{} and wait}
	  \blockif{if \booloperator{\ovalsensing{answer} = \ovalnum{}} then}{\blockspace[0.5]}
    \end{scratch}
  \end{minipage}
  \hfill
  \begin{minipage}{0.7\columnwidth}
    \raisebox{-3mm}{
      \begin{scratch}[scale=0.4]\blocksensing{ask \ovalnum{} and wait}\end{scratch} $\rightarrow$
    }
    \hspace*{-2mm}
    \begin{scratch}[scale=0.4]\blockif{if \boolempty[2em] then}{\blockspace[0.5]}\end{scratch}
    \hspace*{-2mm}
    \raisebox{-3mm}{
      $\rightarrow$
      \setscratch{scale=0.6}\booloperator{\ovalnum{} = \ovalnum{}} $\rightarrow$
      \ovalsensing{answer}\setscratch{scale=0.55}
    }
  \end{minipage}
}

\subfloat[\label{fig:idiom2}Hiding a sprite on change of the backdrop as an isolated script with no other blocks following. The 4-gram has a probability of \textpercent{95.48}.]{
  \begin{minipage}{0.95\columnwidth}
    \textsf{\footnotesize Begin Script} $\rightarrow$
    \begin{scratch}[scale=0.4]\blockinit{When backdrop switches to \selectmenu{backdrop1}}\end{scratch} $\rightarrow$
    \begin{scratch}[scale=0.4]\blocklook{hide}\end{scratch} $\rightarrow$
    \textsf{\footnotesize End Script}
  \end{minipage}
}

\end{center}
\caption{\label{fig:idioms}Common language idioms as captured by the model.}
\end{figure}

Whenever there is a block in the local context that is usually followed by an
expression block, this significantly limits the actual choice of
 matching blocks.
The code example from \cref{fig:example_code:compl} illustrates this phenomenon:
\includegraphics[height=1.1em]{img/repeatuntil} has room only for diamond-shaped expression blocks.
Round expression blocks are intended to go into the round placeholders of the \setscratch{scale=0.6}\booloperator{\ovalnum{} = \ovalnum{}}\setscratch{scale=0.55} block, which reduces the successor blocks.
The top blocks suggested by the model for this scenario are shown in \cref{tab:recommended}.
The model is \textpercent{80.58} confident that a \setscratch{scale=0.7}\ovalvariable{var} should be inserted into the \setscratch{scale=0.6}\booloperator{\ovalnum{} = \ovalnum{}} block.
\ovallook{costume \selectmenu{number}} and \setscratch{scale=0.65}\ovalsensing{answer}\setscratch{scale=0.55} have a comparably low probability of \textpercent{5.69} and \textpercent{5.29} compared to the variable block.
While \ovallook{costume \selectmenu{number}} is a sensible suggestion for this scenario, the \setscratch{scale=0.65}\ovalsensing{answer}\setscratch{scale=0.55} block would only be usable if preceded by a \begin{scratch}[scale=0.4]\blocksensing{ask \ovalnum{} and wait}\end{scratch} block to which the  answer block could refer to (cf.~\cref{fig:idiom} for a usage example), but in \cref{fig:example_code:compl} there is no such block.
Nevertheless, all 3 suggestions do have in common that their shape makes them a syntactically correct building block.
This effect is comparable to the prior observation that syntactic elements
contribute substantially to the predictability of source code~\cite{Rahman2019}.
Accordingly, for categories like pen, sound, or motion, which do not contain
much syntactical constraints, completion performs worse.

Finally, some regularity can be attributed to recurring idioms in common \Scratch code.
Considering n-grams to which the model assigns very high probabilities and excluding known patterns such as starting with a \begin{scratch}[scale=0.4]\blockinit{When \greenflag clicked}\end{scratch} block, we see that the model has learned some idioms that go beyond purely syntactic regularities.
Similar to traditional programming languages, there are certain token sequences that repeat across the boundaries of specific programs.
For example, the 4-gram in \cref{fig:idiom} has a probability of \SI[round-mode=places, round-precision=2]{97.56214937098785}{\percent}, and
%
%
describes requesting a user input and reacting based on a comparison of this.
%
%
While this sequence occurs much less frequently than programs beginning with \begin{scratch}[scale=0.4]\blockinit{When \greenflag clicked}\end{scratch} (7 465 vs. 499 146 occurrences), the model is very confident that \setscratch{scale=0.6}\ovalsensing{answer}\setscratch{scale=0.55} follows as last block for this local context.
This can only be partially explained by occurrence, category, and
shape: Although
\setscratch{scale=0.6}\ovalsensing{answer}\setscratch{scale=0.55} is
an oval block and thus fits syntactically well, there
are numerous other oval blocks in \Scratch. However, the \emph{total}
probability for \emph{all} other blocks of all shapes to follow
this local context is less than \textpercent{2.5}.

Other examples of idioms we observed include isolating functionality
in very short scripts (c.f. \cref{fig:idiom2}) and repetitions of the same
blocks (e.g., inserting elements into a list).
These idioms show a connection of the blocks on a semantic level and therefore
indicate further repetitive, predictable structures apart from pure syntactic
connections, i.e., shapes, and block frequency, i.e.,
occurrences. These idioms are comparable to those discovered in
traditional programming languages.


\summary{RQ1, Completion}{%
  The best model (4-gram) achieves a top-3 accuracy of \accFourThree.
  Prediction quality is influenced by block frequency, shape, and category, but we also found recurring idioms influencing regularities.
}

\subsection{RQ2: Bug Finding with N-gram Models}\label{sec:eval-bugram}

For RQ2 we use the n-gram model in the inverse way compared to RQ1:
Rather than predicting the most likely blocks, we are interested in
the least likely sequences of blocks.

In contrast to the evaluation in the \bugram paper~\cite{Wang2016}, we assume that for the small student solutions we can identify \emph{all} actual errors in the programs using tests.
Thus, in our paper, a test suite acts as ground truth to evaluate whether the most unlikely sequences actually contain erroneous code.
In the original paper, the authors looked for \emph{undetected} bugs and refactoring possibilities in very large projects. Instead of comparing the code with a ground truth like existing tests, it was manually examined for improvement possibilities.

\subsubsection{Experimental Setup}\label{subsec:setup_bugram}

The application scenario of the bug finding task is a programming assignment
given in an educational context. We use the dataset provided with the
replication package of the \whisker~\cite{stahlbauer2019testing} paper on
testing \Scratch programs, consisting of 41 student solutions and one
model solution of a Fruit Catching game (c.f. \cref{img:fruit-catching}).
The objective of the game is to catch as many apples and bananas as possible in 30 seconds by moving
a bowl at the bottom of the screen.
For bananas that touch the ground, the player loses points. The game is lost if an apple drops on the ground.
%
%

We trained a 3-gram model with modified Kneser-Ney smoothing on the model
solution and one student solution which was deemed almost correct using automated tests~\cite{stahlbauer2019testing}.
We use $n = 3$ based on prior results of
Wang et al., who found that
3-gram models perform best in finding bugs and refactoring
opportunities~\cite{Wang2016}.

Using this model, we determined the probabilities for all occurring sequences for
the 41 student programs (that is, including the best student solution).
Intuitively, sequences with lower probability assigned by the model are more
likely to contain bugs.
When extracting sequences we exclude \enquote{loose} code, i.e., blocks and scripts not
connected to an event handler which are never executed.
Since the ideal sequence length for this analysis has not previously been
investigated in our context, we performed the evaluation for sequence lengths
from 3 to 6.
Longer sequence lengths are unlikely to be useful for the generally small
\Scratch programs, as the sequences otherwise would frequently exceed script
boundaries.

To investigate whether low probability sequences indicate bugs, we randomly
selected 10 of the 41 student solutions for manual validation, considering only those with at least 10 sequences, as they are otherwise unlikely to fully implement any aspects of functionality.
For each of these 10 programs, we manually classified the 10 sequences with the
lowest probabilities, for each of the sequence lengths in the range of $3$ to $6$. Two authors independently evaluated whether the
corresponding sequences contained bugs or not.
As ground truth for the existence of bugs we use the extensive \whisker test
suite provided by Stahlbauer et al.~\cite{stahlbauer2019testing} that fully covers the
program behavior. We consider a sequence to contain a bug exactly if it causes the
failure or skipping of one or more test cases. Thus for each sequence, we determined (1) whether the sequence contains at least one bug, and (2) for each failing test whether the sequence contributes to the failure.
In the case of disagreement of the two independent classifications, these
individual cases were discussed again until a consensus was reached. Thus overall, \num{400} sequences were manually classified.

As a baseline, we further selected and classified 10 random sequences per program using the best
sequence length determined by the classification of low
probability sequences, which
allows us to determine if 10 most unlikely sequences are more
likely bugs than random sequences.

%

%
%

\subsubsection{Threats to Validity}\label{subsec:threats_bugram}

Threats to external validity arise as our experiments are based on one task and
student solutions from two school classes, so the results may not generalize to
other tasks or classes. To avoid threats to internal validity, we randomized
the selection of projects to avoid bias. As manual classification may be
influenced by subjective interpretation, each sequence was independently
classified by two authors of this paper to minimize the influence on the
results (inter-rater reliability: \textpercent{88.8}). Furthermore, the same authors classified all sequences to ensure that
the results are consistent and comparable to one another.
To ensure construct validity of our evaluation, we rely on accepted measures
for bug finding, considering the number of buggy sequences as well as
the number of unique bugs.

\subsubsection{Results}\label{subsec:results_bugram}


\begin{table}
\centering
\caption{Percentage of found bugs for each sequence length.}\label{tab:bugram_recall}
\begin{tabular}{lrrrr}
\toprule
Sequence length &      3 &      4 &      5 &      6 \\
\midrule
Bugs found (\%) &  57.58 &  62.88 &  64.39 &  56.06 \\
\bottomrule
\end{tabular}
\end{table}

\begin{table}
\centering
\caption{For sequences of length 4, the bottom 10 most unlikely sequences and 10 random sequences, the Precision@10 for each sequence to contain at least one bug, as well as the proportion of found bugs in total, and the total number of bugs in the program.}\label{tab:bugram_random_bottom_p_bugs}
\begin{tabular}{lrrrrr}
\toprule
{} & \multicolumn{2}{c}{Precision@10} & \multicolumn{2}{c}{\% Bugs Found} & Bugs \\
{} & Bottom & Random &    Bottom & Random & Total \\
\midrule
K6\_S01  &      90.0 &      40.0 &        96.3 &    37.04 &  27.0 \\
K6\_S12  &      60.0 &      90.0 &       80.8 &    26.9 &  26.0 \\
K6\_S15  &      60.0 &      30.0 &        62.5 &    31.3 &  16.0 \\
K6\_S18  &      60.0 &      50.0 &        50.0 &     50.0 &   6.0 \\
K6\_S31  &      30.0 &      10.0 &        75.0 &     12.5 &   8.0 \\
K7\_S03  &      60.0 &      40.0 &       27.3 &    45.5 &  11.0 \\
K7\_S10  &      20.0 &      0.0 &       28.6 &      0.0 &   7.0 \\
K7\_S14  &      30.0 &      20.0 &        60.0 &     20.0 &   5.0 \\
K7\_S17  &      10.0 &      10.0 &        50.0 &     50.0 &   2.0 \\
K7\_S24  &      70.0 &      50.0 &       33.33 &    20.8 &  24.0 \\
\midrule
Average &     49.0 &     34.0 &       62.88 &    28.79 &  13.2 \\
$p$       &    \multicolumn{2}{c}{0.073} &       \multicolumn{2}{c}{0.003} &       \\
$\effectsize$     &     \multicolumn{2}{c}{0.69} &        \multicolumn{2}{c}{0.84} &       \\
\bottomrule
\end{tabular}
\end{table}

\Cref{tab:bugram_recall} lists the overall percentage of bugs found ($b = \#\text{bugs found} / \#\text{bugs in total}$) for different sequence lengths.
Sequences of length 5 find the most bugs overall, closely followed by sequences
of length 4; sequences of length 6 identify the fewest bugs.
The minor difference between sequences of length 4 and 5 originates only from a single program, for which sequence length 5 finds significantly more bugs (K7\_S03: \bugramOutlierFour vs.\ \bugramOutlierFive).
In all other programs, the results of sequence length 4 are equal or even better.
Since sequences of length 4 narrow down the source of the problem/bug better
than sequences of length 5,
%
all further results are based on a sequence length of
4 tokens.


\Cref{tab:bugram_random_bottom_p_bugs} lists the results per program 
based on the least likely and random sequences. Precision@10 is given
in terms of the number of sequences in the bottom 10 containing at least one
bug, i.e., the probability of a sequence in the bottom 10 to contain an actual
bug. The table also shows the proportion of all bugs found in each program.
Since the programs are relatively small, there is frequently more than one bug
per sequence; for reference, the table also lists the total number of bugs,
which corresponds to the number of failed tests.

\looseness=-1
In terms of the percentage of buggy sequences, on average half the
sequences among the least likely ones contain at least one bug, in
contrast to only \textpercent{34} of the randomly selected sequences. The difference
is not significant at $\alpha = 0.05$ (\mbox{$p = 0.073$}), but note
that, since the sequences are randomly selected, there is some overlap
with the 10 least likely sequences: on average \textpercent{16} of the randomly
selected sequences are also among the bottom 10.
However, randomly selected sequences of length 4 are only capable of
finding \textpercent{29} of bugs in total, and they find no more than \textpercent{50} of the
bugs in any program. In contrast, the least likely sequences find an
average of \textpercent{63} of the bugs in total. The improvement of the least
likely sequences over random sequences is statistically significant,
with a large Vargha-Delaney effect size of $\effectsize=0.84$ and $p = 0.003$
using the Mann-Whitney U test.
This confirms that 
indeed the n-gram model
captures the expected structure of the solution.

\begin{figure}
   \centering
   \begin{minipage}[b]{0.41\columnwidth}
        \vspace*{0pt}
        \subfloat[\label{img:fruit-catching}The Fruit Catching game.]{
        \includegraphics[width=\linewidth]{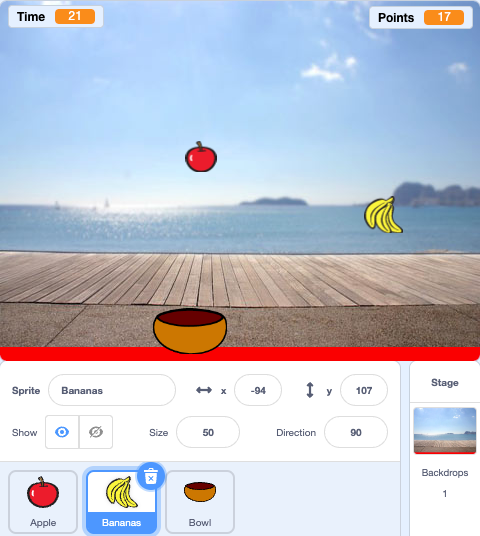}
        }
   \end{minipage}
   \hspace*{2pt}
   \begin{minipage}[t]{0.55\columnwidth}
       \subfloat[\label{img:bugram-bug-example}Student solution K7\_S10 for Banana sprite.
       Colored blocks are examples for reported sequences that cause the failure of tests.]{
        \includegraphics[width=\linewidth]{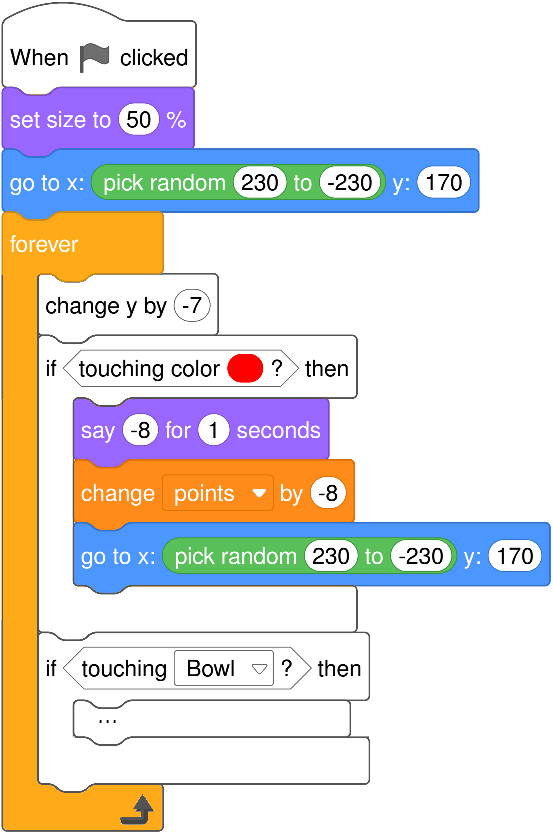}
   }
   \end{minipage}
\caption{\label{fig:fruit-catching}Example for Fruit Catching game with buggy student solution.}
\end{figure}

For example, \cref{img:bugram-bug-example} shows student solution K7\_S10 of the
Banana sprite. The colored parts of the source code are two sequences of length 3 that are among
the 10 least likely according to the model. For both examples, the code actually leads
to failing test cases:
For the first sequence, the solution violates the task specification that
the Banana should start \emph{one second after} the flag was clicked.
The program, however, starts immediately with the for-loop that moves the Banana.
For the second code sequence, the Banana is supposed to \emph{disappear for one second} after the point deduction as a time penalty, but the buggy implementation causes the Banana to start dropping from the top again immediately.



We generally observed 
that blocks not included in the training data have a strong influence on the
probability of a sequence. In particular, not only are such sequences assigned particularly low probabilities, but the same unusual block tends to influence the probability of several surrounding sequences. As a consequence, the least likely 10 sequences may in the worst case cover  only a fraction of the program.
For example, solution K6\_S01 violates the task specification by starting the
game (i.e., each script in each sprite) with pressing the up arrow key instead of the green flag.
The model has not seen this block in the training data. As a result, 9 of 10
sequences reported least likely by the model, contain the unknown
\begin{scratch}[scale=0.4]\blockinit{When \selectmenu{up arrow} key pressed}
\end{scratch}\setscratch{scale=0.55} block.
We noticed the same behavior with uncommon, yet correct, blocks as well.

The original \bugram paper suggests filtering rare tokens~\cite{Wang2016}. Due
to the much more limited vocabulary in \Scratch we decided not to implement
such an approach; however, the experimental results suggest that this could be
a useful addition when implementing this approach in practice, although a
challenge for this will be to not discard too large parts of the rather small
student solutions, which would also cause potential bugs to go undetected.


Program K7\_S17 is the overall best student solution, which was included in the
training data of the model. The fact that one of the two bugs in this program
was identified (\cref{tab:bugram_random_bottom_p_bugs}) shows that, similar to
the \bugram~\cite{Wang2016} approach, errors can even be found
in programs which served as training data.



\summary{RQ2, Bug Finding}{%
For 9 out of 10 programs, the number of bugs found using the model is greater than or equal to that of randomly selected sequences; this improvement is statistically significant.
}


\subsection{RQ3, Comparison: Code Completion with Transformer}\label{sec:eval_transformer}


\subsubsection{Experimental Setup}\label{subsec:setup_transformer}

We trained the RoBERTa model~\cite{liu2019roberta} on a dataset obtained using the same procedure as described in \cref{subsec:setup_compl}, but by sampling \num{500000} programs with \num[round-mode=off]{517431} sprites.
To limit sequences to the maximum length \(m\) that can be processed by the model, we split them into subsequences:
The first generated sequence for a program with \(n\) tokens is \([t_0, \dots, t_{\min(n, m)}]\).
Then the first script not fully included in this sequence is taken as the starting point for the next one.
To embed the script into maximally possible context, tokens preceding and following this script are added symmetrically to fill the sequence up to a length of \(m\).
This is repeated until all tokens have been included in at least one subsequence.
By not splitting sequences at script-level, it is also possible to predict when a script should finish and a new one should be started.
Applying the sequence splitting resulted in \num[round-mode=off]{4493833} sequences for training.
The same \num{10000} programs as for RQ1 were used for evaluation.

We used the default RoBERTa hyperparameters as starting point for further tuning~\cite{liu2019roberta}.
As the language to be modeled is smaller, and programs also tend to be small, the tuning decreased the maximum sequence length to \num{256}, the number of hidden layers (12 to 2), their sizes (hidden size \num{256}, intermediate size \num{512}), and number of attention heads (\mbox{12 to 4}).
The other parameters remained unchanged.


\subsubsection{Threats to Validity}\label{subsec:threats_transformer}

The same threats to validity apply as described in \cref{subsec:threats_completion}.
An additional threat to construct validity arises from the splitting into smaller subsequences. 
To mitigate this risk, we compared the results when splitting the sequence into one padded sub-sequence per script, and using non-overlapping chunks of tokens of the model’s sequence length.
We chose the strategy described in \cref{sec:transformer} as the others did not improve the prediction accuracy.
Our results achieved using the RoBERTa model~\cite{liu2019roberta} might not generalize to alternative transformer-based models.
This matches our aim of not maximizing any particular metric of model performance, but instead providing a baseline and evidence that investigating different models as part of future work is warranted.


\subsubsection{Results}\label{subsec:results_transformer}

\Cref{tab:completion-accuracy} shows the top-\(x\) accuracy of code completion.
The accuracy for the first three suggested tokens is similar to the one for the 3-gram model, i.e., the second best n-gram model according to the results for RQ1.

\Cref{tab:completion-acc-by-group,tab:completion-acc-by-shape} give a more detailed insight which types of blocks can be predicted accurately.
The transformer outperforms the 4-gram model (i.e., the best n-gram model) for predictions of end blocks (\begin{scratch}[scale=0.4]\blockstop{delete this clone}\end{scratch} and \begin{scratch}[scale=0.4]\blockstop{stop \selectmenu{~~~}}\end{scratch}) and performs similarly on regular statements (stack), Boolean expressions (diamond), and placeholders (oval).
For script starts (hat) and branching blocks (c) the accuracy is worse compared to the n-gram model.

The discrepancy between being able to predict ends of scripts and new starts is noteworthy.
Scripts in \Scratch do not have to use an \enquote{end} block as terminal statement.
Therefore, in most cases no clear indicator exists when a new script should start.
Within the \enquote{hat} group of blocks the ones with the best accuracy are
\begin{scratch}[scale=0.4]\blockinit{when backdrop switches to \selectmenu{}}\end{scratch} (\SI{\accTransEventBackdropSwitch}{\percent}),
\begin{scratch}[scale=0.4]\blockinit{when \selectmenu{key} is pressed}\end{scratch} (\SI{\accTransEventKeyPressed}{\percent}), and
\begin{scratch}[scale=0.4]\blockinit{when I receive \selectmenu{}}\end{scratch} (\SI{\accTransEventBroadcaseRecv}{\percent}).
In those cases it is likely that corresponding statements to change the background or send messages are placed near the end of previous scripts, which can then be interpreted as hints that new scripts should start.
Note that the \enquote{end} blocks are not necessarily placed at the end of a sprite sequence.
Instead, the scripts within the sequence are saved in the same order as they were created by the user.
Hence, the model cannot use the number of tokens following the \enquote{end} block to learn if a script should end.

For branching blocks (c-blocks) the transformer has substantially lower accuracy (\printpercent{\accTransShapeC}) compared to the n-gram model (\SI{59.5}{\percent}).
This may be caused by the bidirectional attention that is applied during training:
Using the surrounding context from both sides of the masked c-block, the model learns that they are in most cases, except for forever-loops, followed by a Boolean condition token.
As this information is missing during the code completion task, there is not enough context to reliably predict the correct token.
However, modifying the attention mechanism to only allow unidirectional attention on tokens preceding the masked one resulted in worse accuracy.

Overall, the prediction accuracy of the transformer is worse than the best n-gram model.
Transformers can handle large vocabularies in the order of tens of thousands of different words~\cite{devlin2019bert}, but
this advantage is of little use as the tokenized \Scratch language only has \num{137} words.
Additionally, the usefulness of long range information to the next token prediction is not clear.
The scripts are ordered in the sequence in which the user inserted the first block contained in it while creating the program and do not depend on each other in the program flow except for passed messages in between.
For example, in the context within the same sprite some scripts might handle user inputs and the resulting movement while others are triggered on interactions with other sprites to play sounds or change the look.
Therefore, we expect that the next token mostly depends on the short-range local context and the unrelated blocks of the other scripts act as distracting factor.


\summary{RQ3, Model Comparison}{%
  The transformer model performs comparable to the 3-gram model and thereby worse than our best n-gram model (\(n=4\)).
  We conjecture that this is caused by the small vocabulary and the importance of short-range over long-ranged information for code completion.
}


\section{Discussion}\label{sec:result}

Recent program analyses frequently build on the \enquote{natural hypothesis}~\cite{Hindle2012}, which assumes that programming languages have
similar regularities as natural language, and source code is therefore amenable
to natural language processing techniques. It has been shown that a certain
degree of this observed regularity in source code is due to syntactic overhead
in text-based programming languages. Specifically, Rahman et al.\ showed
lower regularities when filtering common syntax tokens such as delimiters
or nesting tokens~\cite{Rahman2019}. Our
investigation is related in that block-based programming languages explicitly \emph{avoid}
such syntax tokens, thus also making the language potentially less repetitive and predictable.

Compared to \java, the code completion for \Scratch appears
to perform slightly worse. For example, in a related study~\cite{Nguyen2013}
based on a 3-gram model for \java, the \mbox{top-1} accuracy was almost \textpercent{20}
higher than the corresponding \mbox{top-1} accuracy for the 4-gram model in
\Scratch. This study modeled the source code as a stream of lexical
tokens including identifiers, keywords, or symbols, specified by the
programming language~\cite{Nguyen2013}. Further, this study evaluated a
completion task, using local context such as \(<\)\texttt{if, (, node}\(>\) and
completion suggestions such as \enquote{\texttt{!= null}}, \enquote{\texttt{==
null}}, and \enquote{\texttt{.isRoot()}}~\cite{Nguyen2013}. The task design
is comparable to that of our completion task. Accordingly, we assume
that the results are comparable to the extent possible across programming
language boundaries. A difference in prediction accuracy of almost \textpercent{20}
therefore suggests that there are substantial differences either in the
properties of block-based vs.\ text-based programming languages or in the
characteristics of their programs.

Based on recent trends in NLP and software engineering, one might
expect deep learning models to make a difference here, but our results
suggest this is not the case (cf. \cref{subsec:results_transformer}).
In a similar experiment on \java code by Ciniselli et al.\ the
transformer model achieved a better prediction accuracy than their best n-gram
model~\cite{ciniselli2021}. This suggests that there is a difference in how the
model is able to use long-range information in block-based vs.\ text-based
programming languages.
To mitigate this, it may be possible to
integrate more information into a deep learning
model~\cite{guo2020graphcodebert,kim2021}. For example,
using the flat sequences as the input to the transformer removes all
structural information from the code.  Guo et
al.~\cite{guo2020graphcodebert} adapted the attention mechanism of
their GraphCodeBERT model to focus on related code elements determined
by the data flow graph.  Similarly, for \Scratch information about the
relation of scripts (e.g., passed messages, changed sprite attributes)
could be extracted~\cite{Litterbox} to help the transformer focus
on scripts related to the one that should be completed
and ignore other ones. This could improve the capture of long-range
information in a long sequence containing several short scripts.

\begin{figure}[t]
	\centering
	\includegraphics[width=0.99\columnwidth]{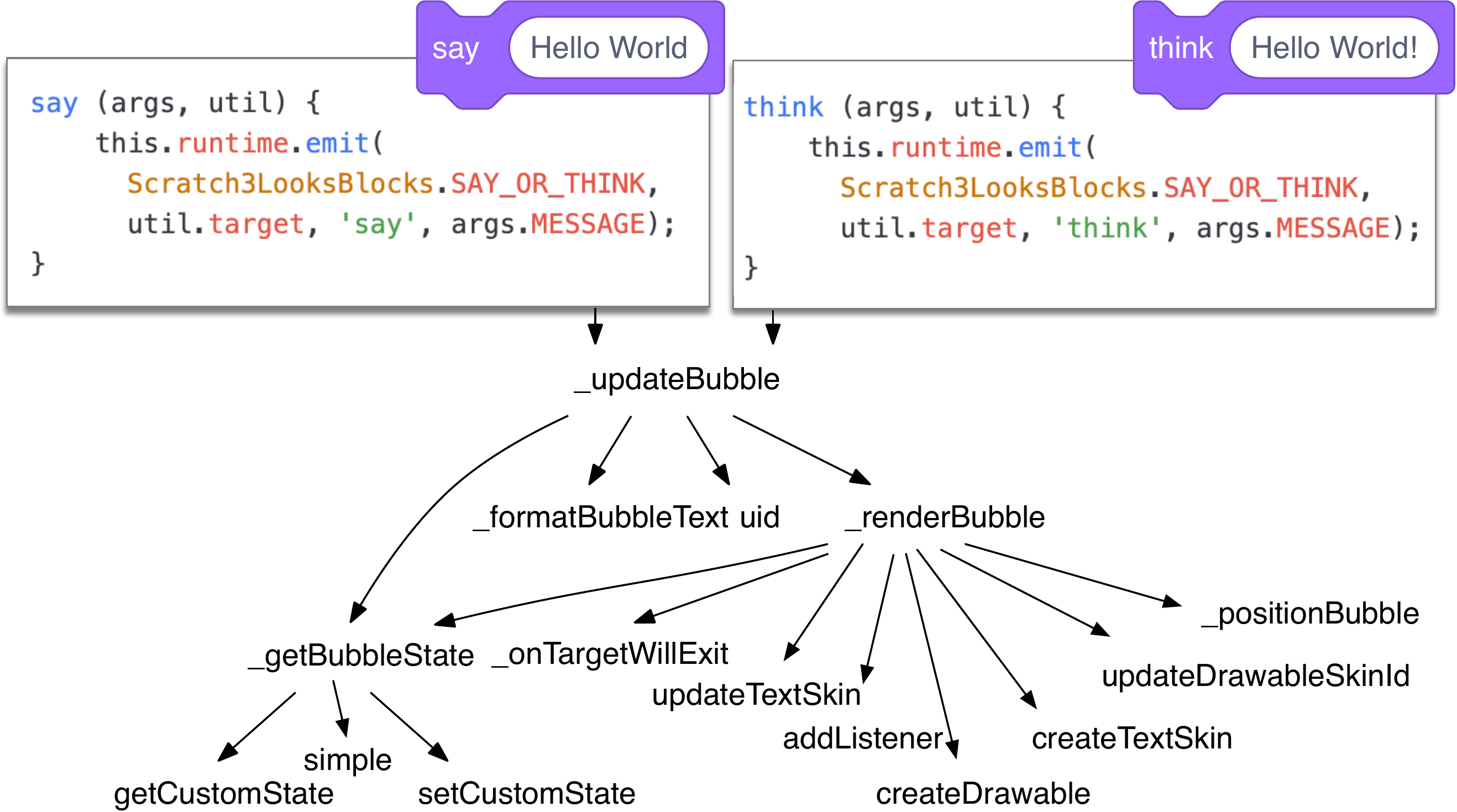}
	\caption{\label{fig:scratch_vs_code}Two different \Scratch blocks vs.\ their shared implementation and call-tree in \javascript.}
\end{figure}

As part of RQ1 and RQ3 (\cref{sec:rq1,sec:eval_transformer}), we discussed some aspects of
block-based programs and their influence on predictability. Besides these
aspects, there is another important difference between traditional, text-based programming
languages and block-based programming languages that likely has an impact on
repetitiveness and predictability: the level of abstraction at which source
code is defined.
\Cref{fig:scratch_vs_code} shows the actual \javascript code that is
executed by a \begin{scratch}[scale=0.4]\blocklook{say \ovalnum{}}\end{scratch} or a \begin{scratch}[scale=0.4]\blocklook{think \ovalnum{}}\end{scratch} block, as well as the corresponding call tree
of the corresponding callback functions in the \Scratch virtual machine code.
On the one hand, the example shows how abstract \Scratch code is compared to
traditional programming languages. Not only is syntactic overhead removed, but to some
extent the code is also streamlined.
On the other hand, this causes repetitions in the \javascript code and
thus may be interpreted as introducing code regularities: in the
\javascript implementation both the \begin{scratch}[scale=0.4]\blocklook{say \ovalnum{}}\end{scratch} and \begin{scratch}[scale=0.4]\blocklook{think \ovalnum{}}\end{scratch} blocks create,
format, and render a speech bubble, and the call to \texttt{\_updateBubble} is
essentially the same for both blocks.
However, in the \Scratch code, the two blocks are distinct without any repetitiveness.
%
%
The \javascript code is thus more repetitive and predictable.

\section{Conclusions}\label{sec:conclusions}

A recent trend in software engineering research is to apply language
models and NLP techniques to text-based programming languages for a
multitude of different tasks. A niche of programming languages
excluded from this trend so far is represented by block-based
languages, which differ from text-based languages by some of the very
properties that make NLP techniques applicable to source code. In
order to shed light on the applicability of language models to
block-based programs, we empirically studied n-gram and transformer
models for \Scratch.  Although our results demonstrate that
block-based languages are more challenging to predict, they
nevertheless demonstrate that the approach is viable.

Prior work on code completion suggests various ways in which our baseline model
could potentially be improved: For example, it is conceivable that other
models, such as statistical graph models~\cite{Rahman2019,Raychev2014} or
neural models (e.g.,~\cite{li2018code,liu2020multi}),
could improve performance. Further filtering of the vocabulary, e.g., to filter
rare blocks~\cite{Wang2016}, might lead to performance
improvements~\cite{babii2019modeling,Karampatsis2020}. The performance could
also be improved by taking additional context into account~\cite{asaduzzaman2014context};
for example, for \Scratch the context could be provided by the sprite or stage
being edited. Similarly, it might be possible to build models for different
types of programs; for example, games might differ fundamentally from animation
or art projects in \Scratch, thus leading to different models.

Besides the performance of the models, an important question for future work
concerns the application of these models. For example, unlike text-based
programming there is no text-cursor at which to display code completion, thus
creating a usability challenge. Nevertheless, the inclusion of a code completion
system into the \Scratch user interface should be feasible, since the blocks are
already presented grouped by their type. A modification of the interface could
show the recommended blocks as such a group.
However, the educational application domain may also suggest
the need for custom models that take the education level into account; for
example, code completion should not recommend blocks that require concepts a
learner is not yet aware of, which could be determined using computational
thinking metrics for \Scratch~\cite{moreno2015dr}. Furthermore, beyond our initial bug finding task, we envision many possible
applications of language models in the educational domain.

To support replication and future work, the source code of the tokenizer and the language models, the datasets, analysis scripts, and the raw results can be found at
\begin{center}
  https://doi.org/10.6084/m9.figshare.19382588.v1
\end{center}


\section*{Acknowledgements}

This work is supported by BMWK (50RM2100B, ``ANUKI''), Bayerische
Forschungsstiftung (AZ-1520-21, ``DeepCode''), and DFG (FR2955/4-1,
``STUNT'').

\balance

\bibliographystyle{IEEEtran}
\bibliography{related}

\end{document}